\documentclass[aps,onecolumn,superscriptaddress,amsmath,showpacs,tightenlines]{revtex4-1}

\usepackage{graphicx}
\usepackage{subfigure}
\usepackage{natbib}
\usepackage{epsfig}
\usepackage{amsfonts}
\usepackage{amsthm,amsmath,amssymb}
\usepackage{mathrsfs}

\usepackage{epstopdf}
\usepackage{xcolor}
\usepackage[toc,page,title,titletoc,header]{appendix}

\usepackage[utf8]{inputenc}

\begin{document}
\title{Entanglement and work statistics in the driven open system}
\author{He Wang}
\affiliation{College of Physics, Jilin University,\\Changchun 130021, China}%
\affiliation{State Key Laboratory of Electroanalytical Chemistry, Changchun Institute of Applied Chemistry,\\Changchun 130021, China.}%
\author{Jin Wang}
\email{jin.wang.1@stonybrook.edu}
\affiliation{Department of Chemistry and of Physics and Astronomy, Stony Brook University, Stony Brook,\\NY 11794-3400, USA}%

\begin{abstract}

We study the entanglement and work statistics in a driven two-qubit system. The regulation of periodic driving has much more versatility and universality in contrast to reservoir engineering in static systems. We found the quasi-steady state entanglement can be amplified effectively by the external drive in certain parameter regimes. The drive extends the range of temperatures or temperature differences at which entanglement can emerge. From the view of the effective Hamiltonian, the addition of the driving alters the inter-qubit coupling and system-bath coupling, which are 
crucial in determining the quasi-steady state. The work statistics are also investigated. The driven system, as a continuous quantum thermal machine, output work continuously and steadily at the quasi-steady state. There is a distinct operation of modes and corresponding performance by changing driving. It can also be understood that the drive changes the effective Hamiltonian, and further the modes of energy exchanges between the system and the baths as well as the work reservoir.

\end{abstract}

\maketitle
\section{Introduction} \label{introduction}

For the past few years, the periodically driven quantum system has drawn much attention. This has been a result of experimental advances in state-of-the-art laser technology, which makes it implementable in the lab. In accordance with the Floquet theorem \cite{Floquet1} and the follow-up theoretical developments \cite{Floquet2,Floquet3}, physical characteristics of the driven system are primarily understood by the so-called effective Hamiltonian, which reflects the periodic driving. Periodic driving largely extends the controllability of the systems since time as a new control dimension is introduced. It gets rid of the down-to-earth difficulty of reservoir engineering in static systems that the parameter is hard to change once the material is prepared. By designing a suitable driving protocol, one can engineer the effective Hamiltonian, which empowers us to have desirable properties and functionalities of the physical systems.  Coherent control through periodic driving (often known as Floquet engineering) has become a commonly used tool in quantum control, which has been used to realize topologically nontrivial systems \cite{topologically1,topologically2,topologically3,topologically4}, nonequilibrium phase transitions \cite{nonequilibrium_phase_transitions1,nonequilibrium_phase_transitions2}, artificial gauge fields \cite{artificial_gauge_fields1,artificial_gauge_fields2,artificial_gauge_fields3}, and discrete time crystals \cite{Floquet_Time_Crystals1,Floquet_Time_Crystals2}. Additionally, it comes into play in the coherent destruction of tunneling \cite{Coherent_destruction_of_tunneling1,Coherent_destruction_of_tunneling2} and the manipulation of spin-orbit coupling \cite{Spin_orbit_coupling1,Spin_orbit_coupling2}, etc. Meanwhile, significant progress has been made toward the experimental realization of small-scale thermal machines where fluctuations play a significant role. The thermal machines in the quantum regime have been realized on several platforms \cite{Spin_Quantum_Heat_Engine,Thermodynamics_in_single_electron_circuits,Tunable_photonic_heat_transport,Nanomechanical_Heat_Engine,Cooling_and_self-oscillation,heat_engine_with_ultracold_atoms,Single-Ion_Heat_Engine,Single-atom_Heat_Engine}. Specific examples include a quantum absorption refrigerator with trapped ions \cite{Quantum_absorption_refrigerator}, quantum heat engines using an ensemble of nitrogen-vacancy centers \cite{Experimental_Heat_Engines}.  Continuous thermal machines \cite{QTM2021} do not require intermittent couplings and decouplings between the working fluid and the baths, which are particularly challenging to implement at microscopic scales, in contrast to their reciprocating counterparts. As a result, they have greater experimental relevance.  Continuous thermal machines are typically implemented by a periodic modulation of the system Hamiltonian, which drives the system to a periodic quasi-steady state in general. The temporal driving methods may drive small systems in nonequilibrium quasi-steady states with far greater versatility and universality than the static manipulation methods, which hinge on steady nonequilibrium sources such as temperature bias, chemical potential difference, etc.\cite{Ren2021}.

A real system is always inevitably influenced by its surroundings and this can lead to the system's decoherence eventually. The environmental effect plays a crucial role in the evolution of the system \cite{The_Theory_of_Open_Quantum_Systems}. How to confront dissipation and decoherence is a fundamental challenge in quantum technology. It has been shown that the dissipation can be effectively suppressed by the formation of the Floquet bound state under temporal driving \cite{Floquet_control_of_quantum_dissipation_in_spin_chains,Floquet_engineering_to_entanglement_protection,Floquet_control_of_quantum_battery}. The scheme to generate a maximally entangled state and then protect it based on Floquet engineering is also proposed \cite{Floquet_engineering_to_entanglement_protection,Floquet_Engineering_Qubits}. However, the entanglement of the nonequilibrium quasi-steady state still lacks of investigations based on our knowledge. Indeed, the balance of the periodic driving and dissipation can yield a variety of nonequilibrium steady states and phase transitions in various systems including cavity-QED systems \cite{cQED1,cQED2}, cold atoms \cite{cold_atoms1,cold_atoms2}, ideal Bose gases \cite{quantum_gases1,quantum_gases2,quantum_gases3}, and so on. A natural question is raised: Is entanglement in the quasi-steady state? Can we enhance it with Floquet engineering? The answer is positive. The entanglement can be magnified significantly by the befitting driving. The system can become entangled in a wider range of temperatures or temperature differences as a result of the drive. From the standpoint of effective Hamiltonian, the existing driving gives rise to the change of inter-qubit coupling and system-bath coupling, as well as further, modifies steady state entanglement. The work statistics of the open system are also of interest. The net flow from the baths to the system disappears when the static system approaches the steady state. When certain external driving is applied, the system can operate continually and steadily as a continuous quantum thermal machine. The different drivings contribute to the various modes of energy exchanges between the system and the baths as well as the work reservoir (energy source of the external agent which modulates the system) in the quasi-steady state, which bring about the thermal machine with  numerous operation of modes. We quantify the performance efficiency of different operation modes, which is bounded by the Carnot limit in general.

The rest of the paper is organized as follows. In Section~\ref{Theoretical framework}, we introduce our model. We then derive a generalized master equation by means of double-projective measurement protocol and Floquet theory. In Section~\ref{entanglement in driven open system}, the quasi-steady state entanglement is studied. In Section~\ref{work statistics in driven open system}, we investigate the work statistics at a quasi-steady state. Finally, we draw our conclusions in Section~\ref{conclusion}. 

\section{Theoretical framework} \label{Theoretical framework}

\begin{figure}[!ht]
    \centering
\includegraphics[width=4in]{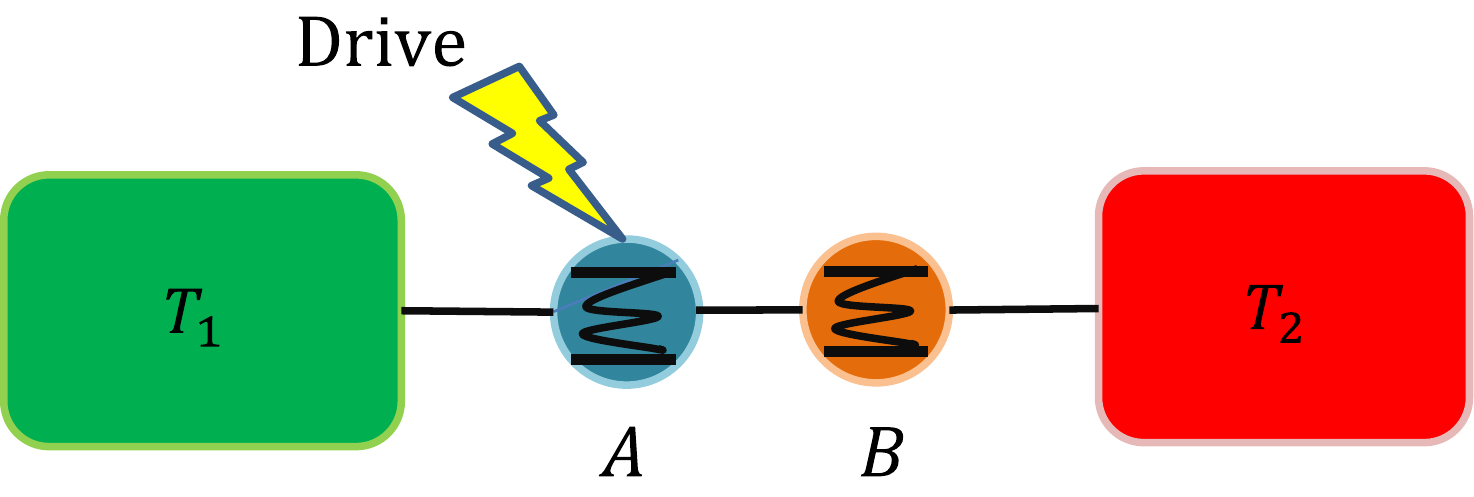}
\caption{\label{fig:0} A sketch of the model. There are two interacting qubits, which are coupled with individual baths as well. The qubit A is driven by the external field. }
\end{figure}

To start, we first introduce the model we studied. The system is composed of a pair of interacting qubits, and each qubit couples to its own bosonic bath with a certain temperature. Meanwhile, the system is driven by external field control. The sketch of the model is shown in Fig.\ref{fig:0}. The Hamiltonian of the whole system is 
\begin{equation}\begin{split}
\label{eq:1}
\pmb H&=\pmb H_{S}(t)+\pmb H_{B}+\pmb H_{SB}\\
&=\sum_{i=A,B}\frac{\omega_{i}+D_{i}(t)}{2}\pmb \sigma_{z}^{i}+\lambda(\pmb\sigma_{+}^{A}\pmb\sigma_{-}^{B}+\pmb\sigma_{-}^{A}\pmb\sigma_{+}^{B})+\sum_{i=A,B}\sum_{k} \omega_{ik}^{2}\pmb a_{ik}^{\dagger}\pmb a_{ik}+ \sum_{i=A,B}\pmb \sigma_{x}^{i} \sum_{k} \pmb c_{ik}\pmb x_{ik},
\end{split}
\end{equation}
where $\pmb \sigma_{z}^{i}$ are Pauli matrices for the i-th qubit, it reads $\pmb \sigma_{z}^{A}=\pmb \sigma_{z}\bigotimes\pmb I$ for the A qubit or $\pmb \sigma_{z}^{B}=\pmb I \bigotimes\pmb \sigma_{z}$ for the B qubit. $\omega_{i}$ is the energy spacing of the i-th qubit. $D_{i}(t)$ is the external field control. $\lambda$ measures the coupling interaction between two qubits. $\pmb a_{ik}^{\dag}(\pmb a_{ik})$ is the creation (annihilation) operator of k-th bosonic mode in i-th bath and satisfies the commute relation $[\pmb a_{ik},\pmb a_{i'k'}^{\dag}]=\delta_{ii'}\delta_{kk'}I$. The $c_{ik}$ are coupling constants that describe the coupling of the i-th qubit to its own reservoir modes $\pmb a_{ik}$. To fully characterize the interaction between the system and baths, we need to define the spectral density of the baths, which follows 

\begin{equation}\begin{split}
J_{i}(\omega)=\frac{\pi}{2}\sum_k \frac{c_{ik}^{2}}{\omega_{ik}}\delta(\omega-\omega_{ik})
\end{split}
\end{equation}
A structure-less spectral density (e.g. linear form $\frac{d_{i}}{\omega_{0}}\omega$) typically empowers a Markovian treatment of the reservoirs due to the fast decay of its associated correlation functions, while a more structured spectral density (e.g. strongly peaked around a frequency $\frac{d_{i}\gamma\omega}{(\omega^2-\omega_{res}^2)^2+\gamma^2\omega^2}$) requires a more involved treatment \cite{Sebastian2018}. In this paper, we simply take a structure-less spectral density $J_{i}(\omega)=\frac{d_{i}}{\omega_{0}}\omega$ into consideration. 

The drives $D_{i}(t)$ may be any time-dependent function, but we only take into account an easily implementable drive scheme: a monochromatic drive with frequency $\omega_{L}$ and amplitude $K$ only acts on the qubit A, i.e., $D_{A}(t)=K\cos{\omega_{L}t}$ and $D_{B}(t)=0$. 

Note that the total Hamiltonian is periodic in time. One can take advantage of the Floquet theorem to solve a time-periodic Schr\"odinger equation $i\partial_t|\psi_r(t)\rangle= \pmb H_S(t)|\psi_r(t)\rangle$, where $\pmb H_S(t) = \pmb H_S(t+T)= \sum_k e^{ i k \,\omega_L t} \pmb H_k$, with period $T=\frac{2 \pi}{\omega_L}$. A solution to this Schr\"odinger equation is given by Floquet states $ |\psi_r(t)\rangle= e^{- i \varepsilon_r t} |r(t)\rangle$, where $\varepsilon_r$ are called quasienergies and $|r(t)\rangle=|r(t+T)\rangle $ are Floquet modes. The existence of Floquet states in time-periodically driven systems follows from the Floquet theorem in a similar way to the existence of Bloch states in spatially periodic systems \cite{Floquet1,Floquet2}.  We also mention that there is a similar Floquet theorem for open systems in \cite{Floquet_theorem_open1,Floquet_theorem_open2}. In Appendix.\ref{Floquet theory}, we review more details on the Floquet theory.

To study the heat statistics of the driven system, we follow the full counting statistics formalism in \cite{Sebastian2018,counting_field1,counting_field2}. The heat moments are expediently described in terms of the characteristic function $G(\chi) = \int d\Delta E \, e^{-i \chi \Delta E } p(\Delta E)$, which is based on double-projective measurement of the environment. With the help of conditional probability $p(E_1;E_0)$, which one measured the energy of environment $E=E_0$ at time $t_0$, and a follow-up measurement gave $E_1$ at time $t_1$, and the probability $p(E_0)$ to measure $E_0$ at time $t_0$, the probability distribution function for the heat energy exchange $\Delta E$ to be transported to the reservoir between times $t= t_0$ and $t= t_1$ can be expressed as

\begin{equation}
p(\Delta E,t) = \sum_{E_1,E_0}\delta(E_1 - E_0 - \Delta E) p(E_1;E_0) p(E_0).
\end{equation} 

Taking account into the projective operator $\pmb P_{E_m}$ and its property $\pmb P_{E_m} \pmb P_{E_n} = \delta_{mn} \pmb P_{E_m}$, we have 
\begin{equation}\begin{split}
p(E_1,E_0)&=Tr[\pmb P_{E_1} \pmb U(t) \pmb P_{E_0} \pmb\rho_{tot}(0)\pmb P_{E_0}\pmb U^{\dagger}(t) \pmb P_{E_1} ]\\
&=Tr[\pmb U^{\dagger}(t) \pmb P_{E_1} \pmb U(t) \pmb P_{E_0} \pmb\rho_{tot}(0)\pmb P_{E_0} ].
\end{split}
\end{equation} 
The generating function is given by 
\begin{equation}\begin{split}
G(\chi,t) &= \int d\Delta E \, e^{-i \chi \Delta E } p(\Delta E) \\
&=\int d\Delta E \, e^{-i \chi \Delta E } \sum_{E_1,E_0}\delta(E_1 - E_0 - \Delta E) p(E_1;E_0) p(E_0)\\
&=\sum_{E_1,E_0} p(E_1;E_0) p(E_0)e^{-i \chi \Delta E } 
\end{split}
\end{equation} 

Assume the initial total density matrix $\pmb\rho_{tot}(0)=\pmb\rho_{S} (0)\otimes \pmb\rho_{B}(0)$ to be factorized into the system density matrix $\pmb \rho_{S} (0)$ and the thermalized environment density matrix $\pmb\rho_{B}(0)= e^{-\beta \pmb H_B} /Z$, where $Z$ is the partition function of the environment.
Particularly, this assumption indicates that all projectors $\pmb P_{E_m}$ commute with $\pmb\rho(0)$, ensuring that the dynamics of the reservoir are unaffected by the initial measurement of the observable. By the use of $e^{-i\chi\pmb H_B} = \sum_{E_m}\pmb P_{E_m}e^{-i\chi E_m}$, we have 

\begin{equation}\begin{split}
G(\chi,t) &=\sum_{E_1,E_0} Tr[\pmb U^{\dagger}(t) \pmb P_{E_1} \pmb U(t) \pmb P_{E_0} \pmb\rho_{tot}(0)\pmb P_{E_0} ]e^{-i \chi \Delta E } \\
&= Tr[\pmb U^{\dagger}(t) \sum_{E_1}\pmb P_{E_1} e^{-i \chi E_{1} }\pmb U(t) \sum_{E_0}\pmb P_{E_0}e^{i \chi E_{0} }\pmb\rho_{tot}(0) ]\\
&= Tr[\pmb U^{\dagger}(t) e^{-i\chi\pmb H_B}\pmb U(t) e^{i\chi\pmb H_B}\pmb\rho_{tot}(0)  ]\\
&= Tr[\pmb U(\chi,t) \pmb \rho_{tot}(0) \pmb U^{\dagger}(-\chi,t) ]\\
&= Tr[ \pmb\rho_{tot}(\chi,t) ],
\end{split}
\end{equation} 
where $\pmb U(\chi,t) = e^{-i\chi\pmb H_B/2}\pmb U(t) e^{i\chi\pmb H_B/2}$ and $\pmb \rho_{tot}(\chi, t) =  \pmb U(\chi,t) \pmb \rho_{tot}(0) \pmb U^{\dagger}(-\chi,t)$. It enables us to compute the statistics of the energy transferred between the system and reservoir by straightforward differentiation 
\begin{equation}
\left\langle \Delta E^n \right\rangle  = - \frac{\partial^n}{\partial (i \chi)^ n} G(\chi) \vert_{\chi=0}.
\end{equation} 


Utilize the modified time evolution operator $\pmb U(\chi,t)$ and modified density matrix $\pmb \rho_{tot}(\chi, t)$, and assume that the coupling between the system and baths are weak enough such that the Born--Markov approximation is valid. We derive a generalized Markov master equation without secular approximation 

\begin{equation} \label{gme}
 \begin{aligned}
\partial_t \pmb\rho(\chi, t) &= \pmb{\mathcal{L}}(\chi,t) \pmb\rho(\chi,t)   \\
 &=-  i \left[   \pmb H_S(t), \pmb \rho(\chi,t) \right]   \\ 
& - \sum_{\left\lbrace i=1,2;\omega;n\right\rbrace } e^{i \Delta_{\omega,n} t }  \lbrace   J_{i}(\Delta_{\omega,n}) N_{i}(\Delta_{\omega,n}) \pmb S_{i,\omega,n}(t) \pmb S_{i}  \pmb \rho(\chi, t) \\
 & -  J_{i}(\Delta_{\omega,n})  \left[1+ N_{i}(\Delta_{\omega,n}) \right]   \pmb S_{i}  \pmb \rho(\chi, t) \pmb S_{i,\omega,n}(t) e^{i \Delta_{i,\omega,n} \chi}   \\ 
 & -  J_{i}(\Delta_{\omega,n})  N_{i}(\Delta_{\omega,n})\pmb S_{i,\omega,n}(t) \pmb \rho(\chi, t)  \pmb S_{i} e^{-i \Delta_{i,\omega,n} \chi} \\ 
 &  +  J_{i}(\Delta_{\omega,n})  \left[1+ N_{i}(\Delta_{\omega,n}) \right] \pmb \rho(\chi, t) \pmb S_{i,\omega,n}(t) \pmb S_{i}\rbrace, 
 \end{aligned}
\end{equation}
where $\pmb S_{1}=\pmb\sigma^{A}_{x}$ and $\pmb S_{2}=\pmb\sigma^{B}_{x}$, $\Delta_{\omega,n} = \omega + n \omega_L$ and 
$\pmb S_{i,\omega,n}(t) = \left[ \int_0^T \frac{dt}{T}  \langle r(t)| \pmb S_{i} \, e^{-i n \omega_l t} |r'(t)\rangle   \right]\, |r(t)\rangle \langle r'(t)| $ 
such that $\omega = \varepsilon_r - \varepsilon_{r'}$. $\varepsilon_r$ is quasi-energy in the first Brillouin zone and $|r(t)\rangle$ corresponds to Floquet modes. $J_{i}(\omega)$ and $N_i(\omega)$ are the spectral density and Bose distribution for the i-th baths respectively. The details are in Appendix.\ref{GME}. Similar master equations have been derived in different research backgrounds \cite{Sebastian2018,counting_field2,Silaev2014,Nafari2022}.

\section{entanglement in the driven open system}\label{entanglement in driven open system}
In this section, we will study the entanglement in the quasi-steady state of the driven open system. Note that due to the periodicity of the Floquet modes $|r(t)\rangle$, the superoperator $\pmb{\mathcal{L}}(\chi,t)$ also has the same periodicity. The evolution of the system can be computed just by setting $\chi = 0$ in Eqn.~(\ref{gme}),
\begin{equation} 
 \begin{aligned}
\partial_t \pmb\rho(t) = &- \, i \left[ \pmb H(t), \pmb \rho(t)    \right] - \sum_{\left\lbrace i=1,2;\omega;n\right\rbrace } e^{i n\omega_L   t }   J_i(\Delta_{\omega,n})  \lbrace     N_i(\Delta_{\omega,n}) 
\left[    \pmb S_i\pmb S_{i,\omega,n}(t) \pmb\rho(t)     -  \pmb S_{i,\omega,n}(t) \pmb \rho(t) \pmb S_i  \right] \\
& + \left[1+ N_i(\Delta_{i,\omega,n}) \right]   \left[    \pmb \rho(t)\pmb S_{i,\omega,n}(t) \pmb S_i   - \pmb S_i \pmb \rho(t)\pmb S_{i,\omega,n}(t)     \right]  \rbrace . 
 \end{aligned}
\end{equation} 
Assuming that in the long-time limit the density matrix $\pmb\rho(t)$ is time-periodic with the same period as the Floquet modes, in the extended space this equation has the form
\begin{equation}
\left[ \sum_k \pmb T_k \otimes  \pmb{\mathcal{L}}_k - i \omega_L \pmb F_z  \right] \vec{\pmb \rho} = 0 \,,
\end{equation}
with $\vec{\pmb\rho} $ a vector containing all Fourier components of $\pmb \rho(t)$. The definition of $\pmb T_k$ and $\pmb F_z$ can be found in Appendix.\ref{Floquet theory}. Therefore, one can numerically obtain a quasi-steady state by truncating the basis of the temporal space. The number of basis should be as large enough as possible to ensure that the result converges.  For general two qubits state $\pmb\rho$, the entanglement can be quantified using the concurrence, which is defined as $C=\max\left\lbrace 0, \lambda_1-\lambda_2-\lambda_3-\lambda_4\right\rbrace$ in bare basis, where $\left\lbrace\lambda_1, \lambda_2, \lambda_3, \lambda_4\right\rbrace$ are the square roots of the eigenvalues of $\sqrt{\pmb\rho}(\pmb\sigma_y \otimes\pmb\sigma_y)\pmb\rho^{*}(\pmb\sigma_y \otimes\pmb\sigma_y)\sqrt{\pmb\rho}$  sorted in a descending order \cite{Quantum_entanglement}.

\begin{figure}[ht]
\includegraphics[width=18cm]{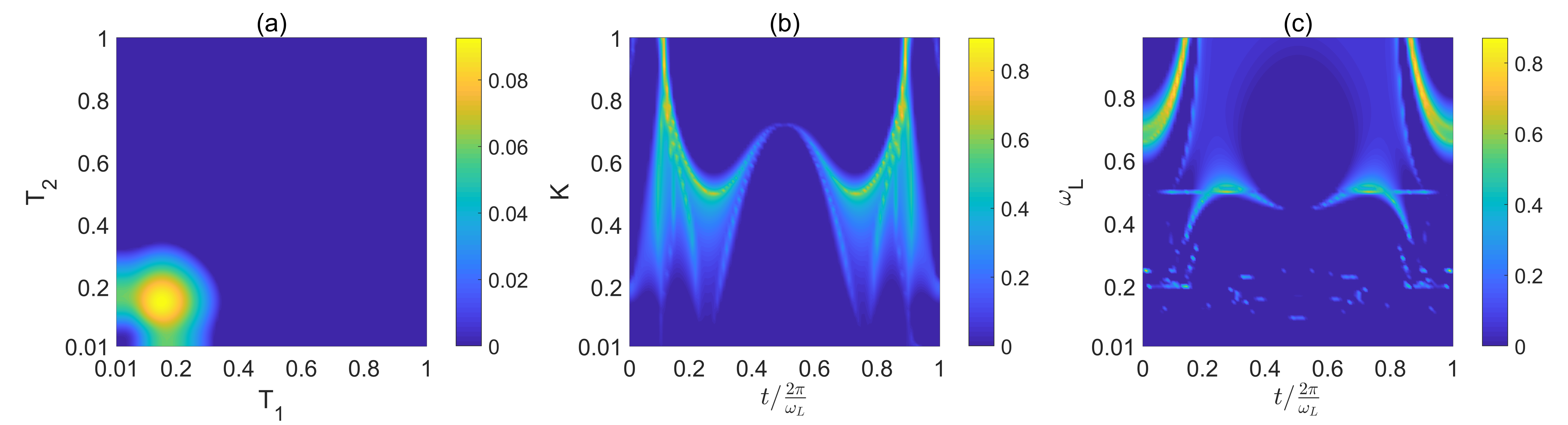}
\caption{\label{fig:1} Entanglement varies w.r.t. (a) the temperatures of baths, (b) the driving amplitude $K$ in one period, and (c) the driving frequency $\omega_{L}$ in one period. The reservoir temperatures are $T_1=0.5$ and $T_2=0.1$ in (a) and (b). And the driving frequency $\omega_{L}=0.5$  for (b) and the driving amplitude $K=0.5$ for (c). Other parameters are $d_{1}=d_{2}=0.001$, $\omega_{01}=\omega_{02}=1$. The coupling strength between two qubits is $\lambda=0.25$, and the energy gaps of the qubits are $\omega_{1}=\omega_{2}=0.5$.}
\end{figure}

We first consider what happened provided that $K=0$, i.e., there is no external driving. This is the case shown in Fig.\ref{fig:1}(a), which comprises both equilibrium and nonequilibrium scenarios. The steady-state entanglement only survives in a limited temperature range. The steady-state entanglement varies non-monotonically with both temperatures or temperature differences. With rising temperatures or temperature differences, the entanglement grows, then diminishes, and finally vanishes. A similar phenomenon has been found in \cite{Wu2019}, which can be phenomenologically explained by the competition between coherence and populations. The system has the same period as the external field when it is driven by an external field.  
The equilibrium/nonequilibrium steady state is replaced with a nonequilibrium quasi-steady state. The entanglement of the nonequilibrium quasi-steady state changes non-monotonically with the driving amplitude. If the driving amplitude is low, entanglement does not exist. If the amplitude is tuned higher, one may be able to harvest more entanglement at certain time slices.  Particularly, the entanglement covers the most time in one period when $K\sim 0.5$ as shown in Fig.\ref{fig:1}(b). The entanglement of the nonequilibrium quasi-steady state also varies non-monotonically with the driving frequency in Fig.\ref{fig:1}(c). The entanglement disappears if the driving frequency is less. One may harvest larger entanglement at some time slices if one tunes the driving frequency properly.  The entanglement lasts longest in one period when $\omega_{L}\sim 0.5$, which is resonant with the qubit.

\begin{figure}[h]
\includegraphics[width=18cm]{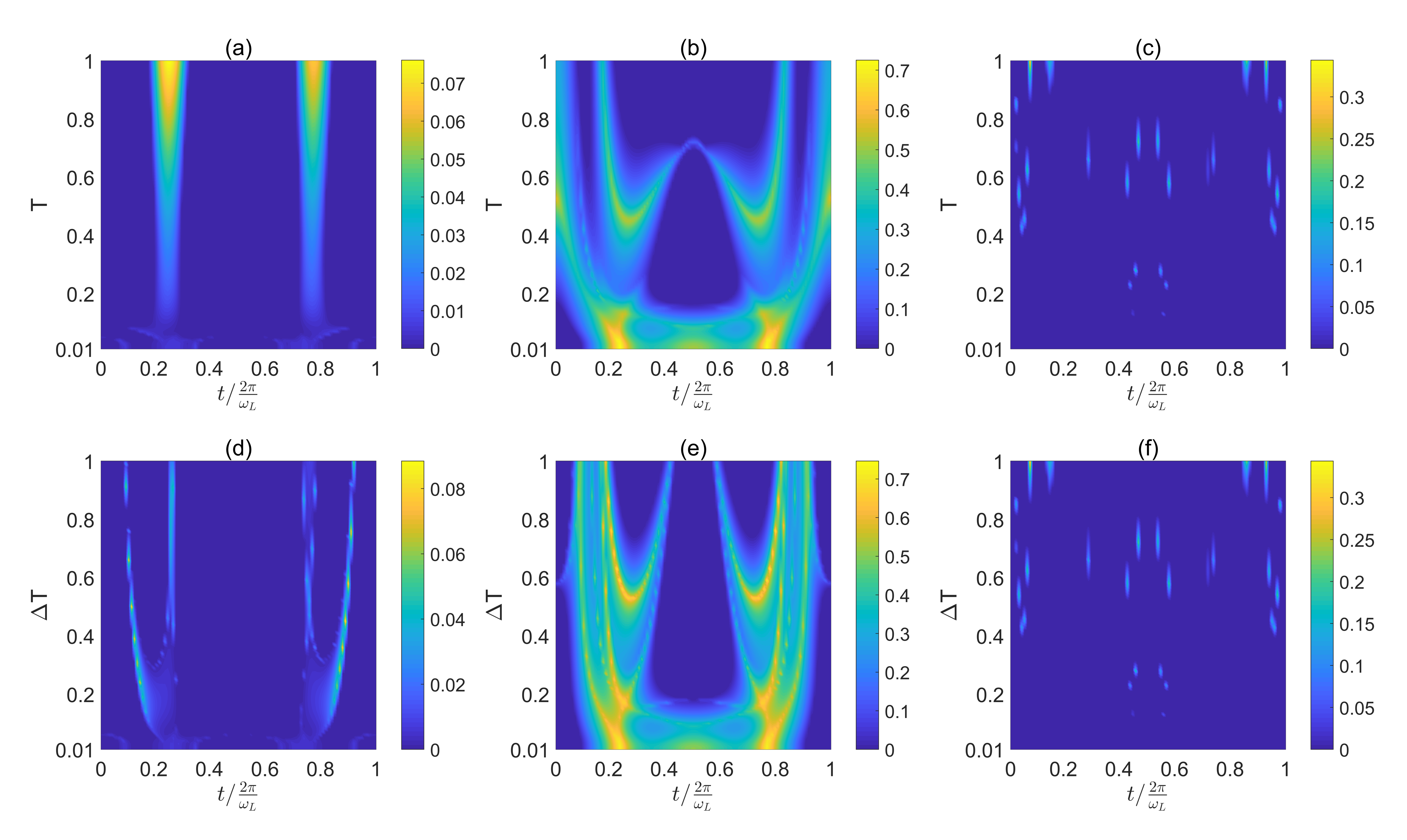}
\caption{\label{fig:2} Variation of entanglement w.r.t temperatures in (a $\sim$ c) and temperature differences in (d$\sim$f) with the driving frequency $\omega_{L}=0.5$, where $K=0.05, 0.5, 5$ for $(a,d), (b,e), (c,f)$ respectively.  $\Delta T=T_1-T_2$ and $T_1$ increases from 0.01. Other parameters are the same as Fig.\ref{fig:1}. }
\end{figure}

Indeed, the nonequilibrium quasi-steady state is a result of the competition between periodic driving and dissipation. Provided that the drive is weak, the dissipation dominates the quasi-steady state. But the quasi-steady state has periodicity as well. This case is depicted in Fig.\ref{fig:2}(a,d). Despite being weak, driving still changes the system significantly. Entanglement even survives at high temperatures or high-temperature differences. Entanglement is visibly enhanced and lasts for the majority of one period if the driving amplitude is comparative with the energy gap of the qubit in Fig.\ref{fig:2}(b,e). Too dramatic driving can also be damaging to the entanglement of the quasi-steady state, as seen in Fig.\ref{fig:2}(c,f), where the driving dominates the quasi-steady state. The behavior of the entanglement is sensitive to temperature changes or variances, therefore the influence of the baths is still considerable.

\begin{figure}[ht]
\includegraphics[width=18cm]{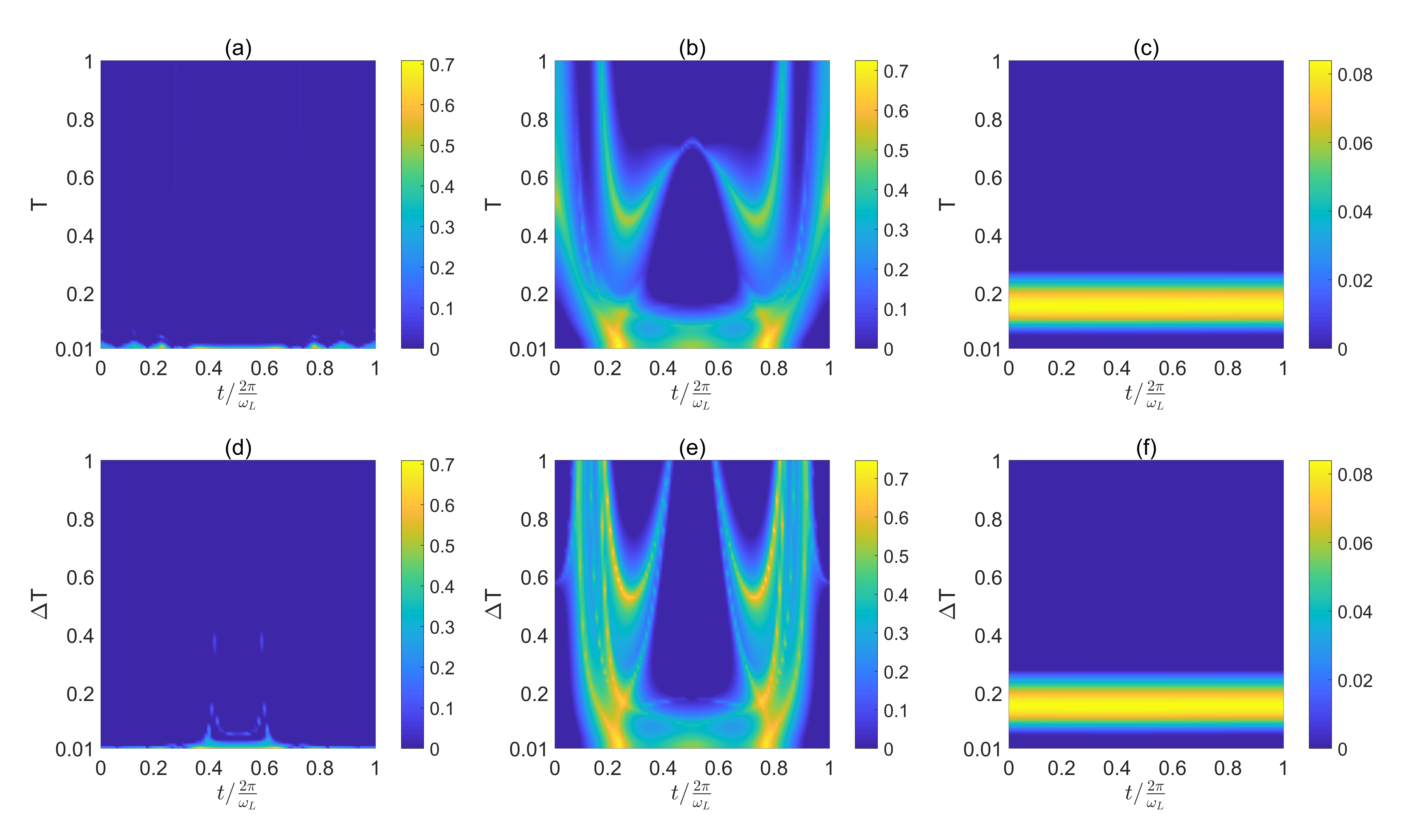}
\caption{\label{fig:3} Variation of entanglement w.r.t temperatures in (a $\sim$ c) and temperature differences in (d$\sim$f) with the driving amplitude $K=0.5$, where $\omega_{L}=0.05, 0.5, 5$ for $(a,d), (b,e), (c,f)$ respectively. $\Delta T=T_1-T_2$ and $T_1$ increases from 0.01. Other parameters are the same as Fig.\ref{fig:1}.  }
\end{figure}

Similar behaviors also happen when we choose different driving frequencies. A driving frequency that is too strong or too weak will be not helpful to harvest more entanglement. When the frequency is low, the harvesting entanglement is considerable only at relatively low temperatures or temperature differences depicted in Fig.\ref{fig:3}(a,d).  Under the Floquet-Magnus expansion approximation \cite{High_frequency_expansion,High_frequency_expansion2}, the effective system Hamiltonian is $\pmb H_{eff}\approx \pmb H_{0}+ \sum_{n>0} \frac{[\pmb H_{n}, \pmb H_{-n}]}{n\omega_{L}}+ \mathcal{O}(\omega_{L}^2)$ in high frequency regime, where $\pmb H_{n}$ is the n-th Fourier component of the system Hamiltonian. One can verify that $\pmb H_{eff}\approx \pmb H_{0}$ in our setting. As a result, when the driving frequency is so high that the system has no time to respond to the external driving, the quasi-steady state entanglement behavior is very similar to that of steady state without driving, as shown in Fig.\ref{fig:3}(c,f). When the frequency is in the middle value, the drive is quite effective at improving the entanglement shown in Fig.\ref{fig:3}(b,e).  

The behavior in the low and medium frequency regime can also be somewhat understood by the effective Hamiltonian. Using the flow equation approach \cite{Floquet3}, we obtain the effective system plus system-reservoirs interaction Hamiltonian (rather than the sole system Hamiltonian) 

\begin{equation}\label{effH}
\pmb H_{eff}^{S}+\pmb H_{eff}^{SB}= \frac{\omega_{A}}{2}\pmb\sigma_{z}^{A}+\frac{\omega_{B}}{2}\pmb\sigma_{z}^{B}+\lambda  J_0( K/\omega_L) (\pmb\sigma_{+}^{1}\pmb\sigma_{-}^{2}+\pmb\sigma_{-}^{1}\pmb\sigma_{+}^{2})+J_{0}( K/\omega_L)\pmb \sigma_{x}^{A} \pmb B_A+\pmb \sigma_{x}^{B} \pmb B_B,
\end{equation}

where $J_0(x)$ is the first kind Bessel function and $\pmb B_i=\sum_K c_{ik}\pmb x_{ik}$ are the reservoir coupling operators. The derivation details of Eqn. \ref{effH} is in Appendix.\ref{flow equation}. The drive affects both the system-reservoirs coupling and the inter-qubit coupling. Note that $J_{0}(0)=1$, we may conclude that the effective Hamiltonian is equal to the original Hamiltonian when the frequency $\omega_{L}$ is set to a very high value and the amplitude $K$ is fixed at a certain finite value, which conforms with the conclusion we got above. Similar to this, when the frequency $\omega_{L}$ is set to a finite value and $K$ is made to be extremely small, the effective Hamiltonian is identical to the original Hamiltonian as well. Therefore, the effect of the drive is not remarkable in the two kinds of limits. This, however, is not the story for $K/\omega_{L}$ being finite. First, the change of the inter-qubit coupling modifies the eigenenergy of the bare system, resulting in an effect on the population distribution at the steady state. Second, the change of system-reservoirs coupling alters the dissipative effect of the environment and further changes the steady coherence. $|J_{0}(s)|\leq1$ indicates that the dissipative effect of one of the baths is weakened and the inter-qubit interaction that induces entanglement within the system diminishes as well.  The end result is generated by the combination of these two effects.  However, we point out that the effective Hamiltonian is unable to provide a master equation to propagate the system. The reason is also clear: the effective Hamiltonian can only describe the stroboscopic time evolution, it can not support a continuous time master equation. However, as we have seen, the effective Hamiltonian still offers certain perspectives for comprehending evolution.

\section{work statistics in the driven open system}\label{work statistics in driven open system}

We investigate work statistics of the driven open system in this section. Using the counting field and its responding generalized master equation, the heat flow from the ith bath to the system is obtained by 
\begin{equation} 
 \begin{aligned}
\dot{Q}_i (t)&= -\partial_t \left\langle \pmb H_{B_i} \right\rangle  = \text{Tr}\left\lbrace \left(  \partial_{i\chi}  \pmb{\mathcal{L}}_{i}(\chi, t) \right)  \cdot \pmb \rho(\chi, t)     \right\rbrace \vert_{\chi=0}  \\
&= \sum_{\left\lbrace \omega;n\right\rbrace } e^{i n\omega_L t }  \quad  \text{Tr}    \lbrace  -  J_i(\Delta_{\omega,n})  \left[1+ N_i(\Delta_{\omega,n}) \right]\Delta_{\omega,n} \pmb S_i \pmb \rho(t) \pmb S_{i,\omega,n}(t) \\ 
& \qquad + J_i(\Delta_{\omega,n})  N_i(\Delta_{\omega,n}) \Delta_{\omega,n} \pmb S_{i,\omega,n}(t) \pmb \rho(t) \pmb S_i \rbrace.
 \end{aligned} 
\end{equation}

\begin{figure}[!h]
\includegraphics[width=14cm]{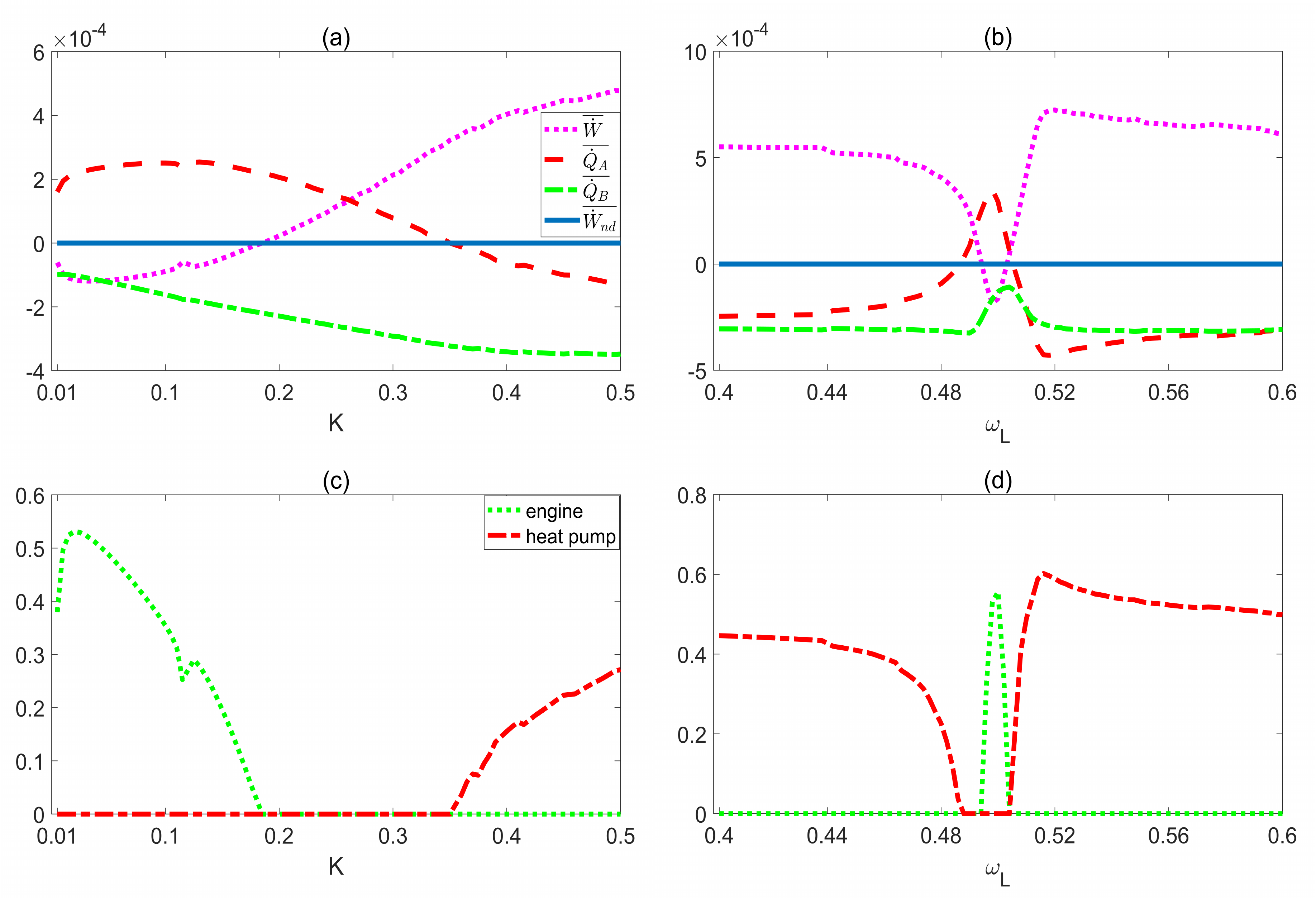}
\caption{\label{fig:5} Variation of mean heat flow and work flow w.r.t the driving amplitude $K$ in (a) and w.r.t. the driving frequency $\omega_{L}$ in (b).  Variation of performance of operation modes w.r.t the driving amplitude $K$ in (c) and w.r.t. the driving frequency $\omega_{L}$ in (d). $\omega_{L}=0.5$ in (a,c), and $K=0.05$ in (b,d). $T_A=1$ and $T_B=0.01$. Other parameters are the same as Fig.\ref{fig:1}. }
\end{figure}

The heat flow, we define, is positive if it flows from the bath to the system. The system must abide by the energy balance at the quasi-steady state, which means $\overline{\dot{ Q}_A}+\overline{\dot{ Q}_B}+\overline{\dot{W}}$=0. In Fig.\ref{fig:5}, we plot the variations of the heat flow and workflow with respect to the amplitude $K$ in (a) and to the frequency in (b). If there is no drive, the system relaxes to the nonequilibrium steady state, and the net flow from the reservoirs to the system vanishes, which also hints that extracted work $\overline{\dot{W}_{nd}}=0$. By varying the driving amplitude, the heat flow from bath B is always negative while the heat flow from bath-A drops from a specific positive value to a certain negative value. The combination of the change of $\overline{\dot{ Q}_A}$ and $\overline{\dot{ Q}_B}$ determines the $\overline{\dot{W}}$. As a result, one may classify thermal operation regimes into three modes in $K$ parameter space in Fig.\ref{fig:5}(a). When $K\in(0.01,0.18)$ roughly, $\overline{\dot{Q}_A}>0, \overline{\dot{ Q}_B}<0$, and $\overline{\dot{W}}<0$, the system operates as a heat engine. Additionally, there is a transition region, $K\in(0.18,0.35)$ approximately, $\overline{\dot{ Q}_A}>0, \overline{\dot{ Q}_B}<0$, and $\overline{\dot{W}}>0$, this operation mode of the system serves no useful purpose. When $K>0.35$, the system functions as a heat pump. Similarly, there are three regimes of thermal operations in $\omega_{L}$ parameter space in Fig.\ref{fig:5}(b). There is a small region around $\omega_{L}=0.5$, $\overline{\dot{Q}_A}>0, \overline{\dot{ Q}_B}<0$, and $\overline{\dot{W}}<0$, the system functions as a heat engine. After going through the narrow transition regime, the system operates as a heat pump when  deviates $\omega_{L}=0.5$ significantly. We also quantify the performance of different operation modes. The efficiency of the heat engine is measured by $|\overline{\dot{W}}/\overline{\dot{ Q}_A}| \leq \eta $ and that of heat pump as $|\overline{\dot{ Q}_A}/\overline{\dot{W}}| \leq \eta^{-1} $, where $\eta=1-\frac{T_B}{T_A}=0.99$ is Carnot bound in this case. We can verify that both the engine and heat pump are bounded by the Carnot limit in Fig.\ref{fig:5}(c,d). As the amplitude grows, the engine's efficiency increases before declining. and the heat pump becomes more efficient. The efficiency of both the heat pump and the engine fluctuates non-monotonically as a function of frequency.

\begin{figure}[!h]
\includegraphics[width=16cm]{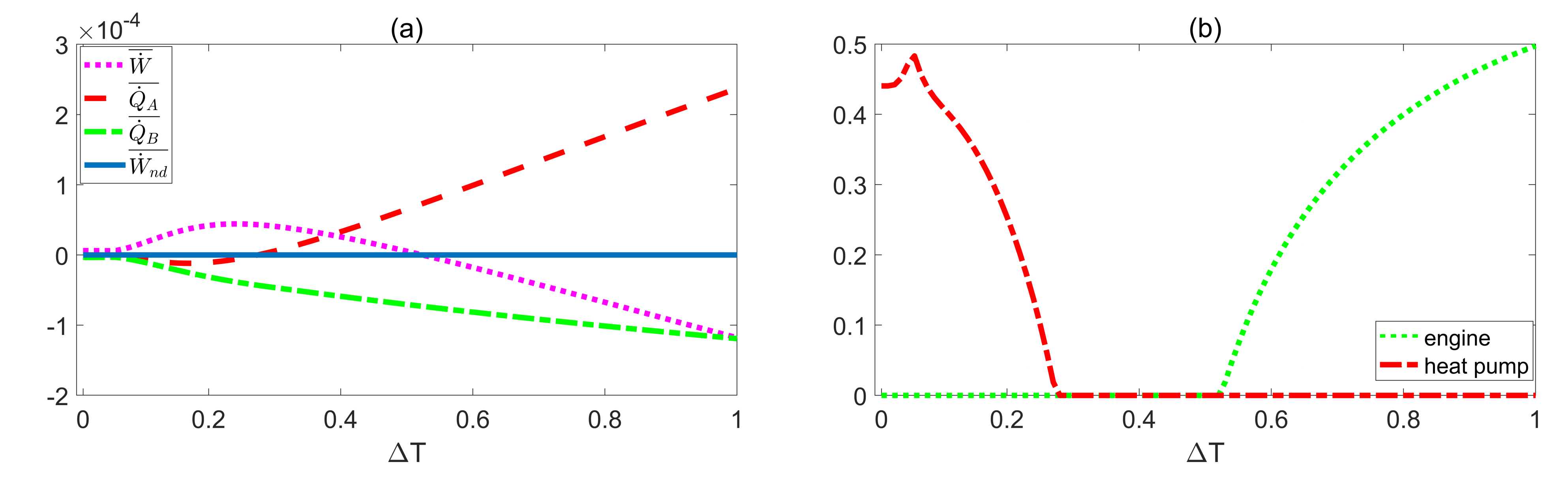}
\caption{\label{fig:6} Variation of mean heat flow and work flow w.r.t the temperature in (a) and Variation of performance of operation modes in (b).  $\omega_{L}=0.5$ and $K=0.05$ in (a,b). $T_A=T_B+\Delta T$ and $T_B=0.01$. Other parameters are the same as Fig.\ref{fig:1}. }
\end{figure}
We can also gain some insights into the reason why the system exhibits distinct operation modes and various performances when we regulate the driving from the perspective of the effective Hamiltonian. The change of inter-qubits coupling and the system-reservoirs coupling re-determines the heat flow and furthermore leads to output work being modified. The heat flow is also changed directly by controlling temperature difference.
If the drive is fixed, the system displays various operation modes as the temperature difference changes. When the temperature difference is less, the system performs as a heat pump; after passing through a transition region, the system behaves as a heat engine at higher temperature differences. This is depicted in Fig.\ref{fig:6}(a). The performance efficiency of the heat pump is lowered as the temperature difference increases, on the contrary, that of the engine increases as illustrated in Fig.\ref{fig:6}(b).

\section{Conclusion}\label{conclusion}
In conclusion, we computed the nonequilibrium quasi-steady state entanglement within the two-qubit system by deriving a generalized master equation. We have better control over the system with external field regulation. The system may harvest more entanglement from the reservoirs compared with the static system. The driven system can be entangled even at high temperatures or temperature differences in contrast to the system without driving. We try to get to the bottom of why the driven system behaves differently from the static system from the effective Hamiltonian. The introduced driving modifies the inter-qubit coupling and system-reservoirs coupling. The total effect influences the quasi-steady state. We also investigate the work statistics in the driven open system. Changing the drive frequency or amplitude, the system will have different modes of operation, e.g., heat pump and engine or others. Their performances related to the heat flow from the reservoirs to the system can be changed by modifying driving and temperature differences. As it stands, our model is likely to be realizable with state-of-the-art laser technology and quantum simulation platforms. Further optimization of the model parameters may relax experimental requirements.

\appendix
\section{Floquet Theory and the Extended Space} \label{Floquet theory}
We give a more detailed introduction to Floquet theory in this section \cite{Floquet2,Sebastian2018}. Consider a time-periodic Hamiltonian $\pmb H_S(t) = \pmb H_S(t+T)= \sum_k e^{ i k \,\omega_L t} \pmb H_k$, with period $T=\frac{2 \pi}{\omega_L}$, according to Floquet theory, a solution to Schr\"odinger equation $i\partial_t|\psi_r(t)\rangle= \pmb H_S(t)|\psi_r(t)\rangle$ is given by Floquet states $ |\psi_r(t)\rangle= e^{- i \varepsilon_r t} |r(t)\rangle$, where $\varepsilon_r$ are dubbed as quasienergies and $|r(t)\rangle=|r(t+T)\rangle $ are Floquet modes (states). Similar to the way Bloch states exist in spatially periodic systems, the existence of Floquet states in time-periodically driven systems comes from the Floquet theorem. \cite{Floquet1,Floquet2}.  We also mention that there is a similar Floquet theorem for open systems in \cite{Floquet_theorem_open1,Floquet_theorem_open2}.

One can ignore the micromotion and focus on the time evolution in a stroboscopic fashion in steps of the driving period $T$ as long as the dynamics we studied over a time span that is long compared to a single driving period. Such a stroboscopic time evolution is governed by the time-independent Floquet Hamiltonian $\pmb H^F_{t_0}$, which is defined in a way that it generates the time evolution over one period,
\begin{equation}\label{eq:DefHeff}
\exp\bigg(-iT \pmb H^F_{t_0}\bigg)=\pmb U(t_0+T,t_0).
\end{equation} 
and can be expressed like
\begin{equation}\label{eq:HeffModes}
\pmb H^F_{t_0} =\sum_r \varepsilon_r |r(t_0)\rangle\langle r(t_0)| .
\end{equation} 
The parametric dependence on the initial time $t_0$ is periodic, $\pmb H^F_{t_0+T}=\pmb H^F_{t_0}$, and related to the micromotion.  One can construct a Floquet Hamiltonian for a different initial time $t_0'$ by applying a unitary transformation, $ \pmb H^F_{t_0'}= \pmb U^\dag(t_0,t_0') \pmb H^F_{t_0}\pmb U(t_0,t_0')$, on a Floquet Hamiltonian $ \pmb H^F_{t_0}$ obtained for the initial time $t_0$. It is difficult to obtain the Floquet Hamiltonian in general, however, various methods are developed based on high-frequency expansion \cite{Floquet2,High_frequency_expansion,High_frequency_expansion2} or others \cite{Floquet3}.

We can introduce a unitary operator that describes the periodic time dependence of the Floquet modes, i.e., the micromotion. The corresponding two-point micromotion operator can be defined by
\begin{equation}\label{eq:Umicro}
\pmb P(t_2,t_1) = \sum_r |r(t_2)\rangle\langle r(t_1)|
\end{equation}
as a result of its construction, it is periodic in both arguments, $\pmb P(t_2+T,t_1)=\pmb P(t_2,t_1+T)=\pmb P(t_2,t_1)$, and evolves the Floquet modes in time,
\begin{equation}
|r(t_2)\rangle=\pmb P(t_2,t_1)|r(t_1)\rangle.
\end{equation}

The Floquet Hamiltonian and the micromotion operator can be written down immediately using Eqn.\ref{eq:HeffModes} and Eqn.\ref{eq:Umicro} if the Floquet states and their quasienergies are known by diagonalizing the time evolution operator over one period. One can then write out the time evolution operator using the Floquet Hamiltonian and the micromotion operator as

\begin{equation}\label{eq:Evolution}
\pmb U(t_2,t_1)
=e^{-i(t_2-t_1) \pmb H^F_{t_2}}\pmb P(t_2,t_1)
=\pmb P(t_2,t_1) e^{-i(t_2-t_1) \pmb H^F_{t_1}}.
\end{equation}

From the above analysis, we can see that quasienergies and Floquet modes are extremely crucial for the evolution of the driven system. We show how to solve them numerically in the following.
The Floquet modes are time-periodic and form a complete basis. To find them one solves the eigenvalue problem
\begin{equation}\label{Flo_eigenvalue}
\left(  \pmb H_S(t) - i \partial_t       \right)  |r(t)\rangle= \varepsilon_r |r(t)\rangle.
\end{equation}

The periodicity of the Floquet modes enables us to map the eigenvalue problem to a time-independent one in an extended Hilbert space, also known as Sambe space \cite{Sambe_space}. To do this, an infinite-dimensional space with integer quantum numbers is introduced. Its basis is given by
\begin{equation}
\mathscr{H}_{T} = \left\lbrace \ldots, |-3\rangle ,|-2\rangle  ,|-1\rangle  ,|0\rangle ,|1\rangle  ,|2\rangle, |3\rangle            \ldots      \right\rbrace. 
\end{equation}

There are two operators $ \pmb T_k$ and $\pmb F_z$ in Sambe space, which are defined as 
\begin{equation}\label{tempo_operator}
\pmb T_k|m\rangle = |m+k\rangle, \qquad \pmb F_z |m\rangle = m |m\rangle. 
\end{equation}

Combining the basis denoted by $\mathscr{H}_{S} = \left\lbrace |\phi_1\rangle ,|\phi_2\rangle ,|\phi_3\rangle \ldots |\phi_K\rangle \right\rbrace$ for the quantum system living in original Hilbert space $\mathscr{H}_S$, one can construt the basis of the extended Hilbert space,
\begin{equation}
\mathscr{H} =  \mathscr{H}_{T}\otimes \mathscr{H}_{S} = \big\{ |n,\phi \rangle\rangle \, |n\in\mathbb{Z},i\in\{1,\dots,K\}\big\},
\end{equation}
with the scalar product $\langle \langle u|v\rangle\rangle = \frac{1}{T} \int_0^T dt \langle u(t)|v(t)\rangle= \overline{\langle u(t)|v(t)\rangle}$ in the extended space. We have denoted vectors in the Sambe space by a double ket notation $|r\rangle\rangle$, which corresponds to $|r(t)\rangle$ in $\mathscr{H}_{S}$. Moreover, a periodic time-dependent operator $ \pmb O(t)= \sum_k \pmb O_k e^{i \omega_L t}$ in extended space is expressed as
\begin{equation}
\pmb O_{ext} = \sum_k \pmb T_k \otimes \pmb O_k.
\end{equation}

With the above definitions, we rewrite the operator $ \pmb Q = \pmb H(t)-i \partial_t$  and solve eigenvalue problem in extended space as
\begin{equation}
\pmb Q_{ext} =\sum_k \pmb T_k \otimes \pmb H_k + \omega_L \pmb F_z \otimes \pmb 1  \qquad \longrightarrow \qquad \pmb Q_{ext} |r_m\rangle\rangle = \varepsilon_{rm} |r_m\rangle\rangle,
\end{equation}
where $\pmb H_k$ are the Fourier components of the Hamiltonian $\pmb H(t)$.  Note that $\varepsilon_r$ is
periodic with period $\omega_L$ because $| r(t)\rangle e^{i m \omega_L t}$ is also the eigenstate of Eqn.\ref{Flo_eigenvalue} with eigenvalue $\varepsilon_{rm}=\varepsilon_r + m \omega_l$. This is the reason why the eigenstates and quasienergies have been denoted with an additional index $m$ in the extended space. The complete set of solutions of the quasienergies eigenvalue problem Eqn.\ref{Flo_eigenvalue} contains a lot of redundant information. All Floquet states of the system can, thus, be constructed, for example, from those Floquet modes whose quasienergies lie in a single Brillouin zone of the $\omega_L$ periodic quasienergy spectrum.  From now on we will denote Floquet modes in the extended space just by $|r\rangle\rangle$, assuming that all lie in the same Brillouin zone.

One can re-formulate the evolution of any operator with the help of The Floquet modes. Take into account an arbitrary operator $\pmb S$.
\begin{equation}\begin{split}\label{S_decomp}
\pmb S(t) &= \pmb U^{\dagger}(t) \pmb S \pmb U(t)\\
&=(\pmb P(t,0) e^{-it \pmb H^F_{0}})^{\dag} \pmb S \pmb P(t,0) e^{-it \pmb H^F_{0}}\\
&=e^{it\pmb H^F_{0}}\pmb P^{\dag}(t,0)\pmb S \pmb P(t,0) e^{-it \pmb H^F_{0}}\\
&=\sum_{r}e^{it \varepsilon_{r}}|r(0)\rangle\langle r(0)| \sum_{m}|m(0)\rangle\langle m(t)| \pmb S \sum_{r'}|r'(t)\rangle\langle r'(0)| \sum_{m'}e^{-it\varepsilon_{m'}}|m'(0)\rangle\langle m'(0)|\\
&=\sum_{r,r'}e^{it (\varepsilon_{r}-\varepsilon_{r'})}\langle r(t)|\pmb S |r'(t)\rangle |r(0)\rangle\langle r'(0)|\\
&=\sum_{\omega,n}e^{it (\omega+n\omega_{L})}\pmb S_{\omega, n} .
\end{split}
\end{equation}
where $\pmb S_{\omega, n} = \left[ \int_0^T \frac{dt}{T} \langle r(t)| S \, e^{-i n \omega_l t} |r'(t)\rangle  \right]\, |r(0)\rangle \langle r'(0)|$ and $\varepsilon_r - \varepsilon_{r'} = \omega$.  Due to the periodicity of the Floquet modes $|r(t)\rangle (|r'(t)\rangle)$, we can perform the Fourier transform, i.e.  $\langle r(t)|\pmb S |r'(t)\rangle |r(0)\rangle\langle r'(0)| =\sum_{k} e^{in\omega_{L}t}\pmb S_{\omega,n} $, where $\pmb S_{\omega, n} = \left[ \int_0^T \frac{dt}{T} \langle r(t)| S \, e^{-i n \omega_l t} |r'(t)\rangle  \right]\, |r(0)\rangle \langle r'(0)|$. In extended space, using Eqn.\ref{tempo_operator} this is easily computed by
\begin{equation}
\int_0^T \frac{dt}{T}  \langle r(t)|  \pmb S \, e^{-i n \omega_L t} |r'(t)\rangle = \langle \langle r| \pmb T_{-n}\otimes \pmb S |r\rangle \rangle.
\end{equation}
It is now straightforward to obtain a master equation for a driven open quantum system thanks to decomposition Eqn.\ref{S_decomp}.

\section{The derivation of the Generalized Master Equation} \label{GME}
The total Hamiltonian, including system and reservoir is $ \pmb H_{tot}=\pmb H_S(t) +\pmb H_B +\pmb H_{SB}$ and we assume an initial factorizing state of the form $\pmb \rho_{tot}(0) = \pmb \rho(0) \otimes \pmb \rho_B $, with $\pmb\rho_B \sim e^{-\beta \pmb H_B}$. Define the  modified density matrix

\begin{equation}\label{eq:CF_def}
\pmb \rho_{tot}(\chi, t) =  \pmb U(\chi,t) \pmb \rho_{tot}(0) \pmb U^{\dagger}(-\chi,t) , 
\end{equation}

with total (system plus reservoirs) initial density matrix $\pmb \rho_{tot}(0)$ and modified evolution operator $\pmb U(\chi,t) = e^{-i\chi\pmb H_B/2}\pmb U(t)e^{i\chi\pmb H_B/2}$, where $\pmb U(t)$ is the evolution operator corresponding to Hamiltonian $\pmb H$. The variable $\chi$ is usually referred to as the counting field. The evolution of operator $\pmb \rho_{tot}(\chi, t)$  is given by 
\begin{equation}
\partial_t \pmb \rho_{tot}(\chi, t) = -i \left[  \pmb H(\chi,t) \pmb \rho_{tot}(\chi,t) - \pmb \rho_{tot}(\chi,t) \pmb H(-\chi,t) \right] . \label{partial_total_rho}
\end{equation}

Taking the trace over the reservoir degrees of freedom we define
\begin{equation}
 \pmb \rho(\chi, t) = \text{Tr}_B \left\lbrace \pmb \rho_{tot}(\chi, t) \right\rbrace .
\end{equation} 
Note that the total density matrix and the reduced density matrix of our system are both recovered by setting $\chi=0$. 


Transform Eqn.~\ref{partial_total_rho} into the interaction picture by $\tilde{\pmb A}(t) = \pmb U_0^{\dagger}(t)\pmb A \pmb U_0(t) $, with $\pmb U_0(t)$ the evolution operator associated to Hamiltonian $\pmb H_0(t)=\pmb H_S(t)+\pmb H_B$. And considering the standard Born--Markov approximations \cite{The_Theory_of_Open_Quantum_Systems}, we obtain  
 \begin{equation}
 \begin{aligned}
 \partial_t \tilde{\pmb \rho}(\chi, t) = 
  &- \int^{\infty}_0 ds \text{Tr}_B \lbrace  \tilde{\pmb H}_I(\chi,t) \tilde{\pmb H}_I(\chi,t-s) \tilde{\pmb \rho}(\chi, t) \pmb \rho_B   
 - \tilde{\pmb H}_I(\chi,t) \tilde{\pmb \rho}(\chi, t) \pmb \rho_B \tilde{\pmb H}_I(-\chi,t-s)  \\
 &-   \tilde{\pmb H}_I(\chi,t-s) \tilde{\pmb \rho}(\chi, t) \pmb \rho_B \tilde{\pmb H}_I(-\chi,t)  
 +   \tilde{\pmb \rho}(\chi, t) \pmb \rho_B \tilde{\pmb H}_I(-\chi,t-s) \tilde{\pmb H}_I(-\chi,t)  \rbrace. 
 \end{aligned}
 \end{equation}
 

We take the interaction Hamiltonian with the form $\pmb H_{SB} = \pmb S \otimes \pmb B=\pmb S \otimes \sum_k c_k \pmb x_k$ and define the correlation function $C(\chi, t) \equiv \langle \tilde{\pmb B}(\chi,t)\pmb B \rangle = \text{Tr}_B\left\lbrace   \tilde{\pmb B}(\chi,t)\pmb B  \pmb \rho_B \right\rbrace $.  Using the fact that $\left\langle \tilde{\pmb B}(\chi,t) \tilde{\pmb B}(\xi, s)    \right\rangle  = \left\langle \tilde{\pmb B}(\chi-\xi,t-s) \pmb B \right\rangle$, we have
 \begin{equation}
  \begin{aligned}
 \partial_t \tilde{\pmb \rho}(\chi, t) = 
  &- \int^{\infty}_0 ds  \lbrace  C(0,s)  \tilde{\pmb S}(t)\tilde{\pmb S}(t-s) \tilde{\pmb \rho}(\chi, t)   
 -      C(-2\chi,-s)  \tilde{\pmb S}(t)  \tilde{\pmb \rho}(\chi, t) \tilde{\pmb S}(t-s) \\
 &-   C(-2\chi,s)   \tilde{\pmb S}(t-s) \tilde{\pmb \rho}(\chi, t)  \tilde{\pmb S}(t)
 +  C(0,-s)   \tilde{\pmb \rho}(\chi, t) \tilde{\pmb S}(t-s) \tilde{\pmb S}(t)\rbrace. 
  \end{aligned}
 \end{equation} 

The correlation functions can actually be written as 


\begin{equation}\begin{split}
C(-2\chi,t) =& Tr_B( e^{i\chi\pmb H_B} e^{it\pmb H_B}\pmb B e^{-it\pmb H_B} e^{-i\chi\pmb H_B} \pmb B \pmb \rho_B )\\
=&Tr_B( e^{i\chi\pmb H_B} e^{it\pmb H_B} \sum_k c_k \pmb x_k e^{-it\pmb H_B} e^{-i\chi\pmb H_B} \sum_{k'} c_k' \pmb x_k' \pmb \rho_B )\\
=&Tr_B( e^{i\chi\pmb H_B} e^{it\pmb H_B} \sum_k c_k \sqrt{\frac{\hbar}{2m\omega}}(\hat{\pmb a}_k+\hat{\pmb a}^{\dag}_k) e^{-it\pmb H_B} e^{-i\chi\pmb H_B} \sum_{k'} c_k' \sqrt{\frac{\hbar}{2m\omega}}(\hat{\pmb a}_{k'}+\hat{\pmb a}^{\dag}_{k'}) \pmb \rho_B )\\
\end{split}
\end{equation}
Set $\hbar=m=1$ and use $\hat{\pmb a}^{\dag}_{k}(t)=e^{it\pmb H_B}\hat{\pmb a}^{\dag}_{k} e^{-it\pmb H_B}=\hat{\pmb a}^{\dag}_{k}e^{iw_kt} $ and $\hat{\pmb a}_{k}(t)=e^{it\pmb H_B}\hat{\pmb a}_{k} e^{-it\pmb H_B}=\hat{\pmb a}_{k}e^{-iw_kt} $. We derive $\langle \hat{\pmb a}_{k}(t)\hat{\pmb a}_{k'}(t')\rangle=\langle \hat{\pmb a}^{\dag}_{k}(t)\hat{\pmb a}^{\dag}_{k'}(t')\rangle=0$, $\langle \hat{\pmb a}^{\dag}_{k}(t)\hat{\pmb a}_{k'}(t')\rangle=\delta_{k,k'}e^{i\omega_k(t-t')}N(\omega_k)$, and $\langle \hat{\pmb a}_{k}(t)\hat{\pmb a}^{\dag}_{k'}(t')\rangle=\delta_{k,k'}e^{-i\omega_k(t-t')}(N(\omega_k)+1)$. Going on the derivation,
\begin{equation}\begin{split}
C(-2\chi,t)=&Tr_B( e^{i\chi\pmb H_B} e^{it\pmb H_B} \sum_k c_k \sqrt{\frac{\hbar}{2m\omega_k}}(\hat{\pmb a}_k+\hat{\pmb a}^{\dag}_k) e^{-it\pmb H_B} e^{-i\chi\pmb H_B} \sum_{k'} c_k' \sqrt{\frac{\hbar}{2m\omega_k}}(\hat{\pmb a}_{k'}+\hat{\pmb a}^{\dag}_{k'}) \pmb \rho_B )\\
&=\sum_{k,k'}\frac{c_k c_k'}{2\omega_k} Tr_B( e^{i\chi\pmb H_B} e^{it\pmb H_B} (\hat{\pmb a}_k+\hat{\pmb a}^{\dag}_k) e^{-it\pmb H_B} e^{-i\chi\pmb H_B} (\hat{\pmb a}_{k'}+\hat{\pmb a}^{\dag}_{k'}) \pmb \rho_B )\\
&=\sum_{k,k'}\frac{c_k c_k'}{2\omega_k} Tr_B( (\hat{\pmb a}_k e^{-i\omega_k(t+\chi)}+\hat{\pmb a}^{\dag}_k e^{i\omega_k(t+\chi)})(\hat{\pmb a}_{k'}+\hat{\pmb a}^{\dag}_{k'}) \pmb \rho_B )\\
&=\sum_{k}\frac{c_k^2}{2\omega_k}  [(N(\omega_k)+1)e^{-i\omega_k(t+\chi)}+ N(\omega_k)e^{i\omega_k(t+\chi)}]\\
&=\int_{0}^{\infty} d\omega J(\omega) [(N(\omega)+1)e^{-i\omega(t+\chi)}+ N(\omega)e^{i\omega(t+\chi)}]\\
\end{split}
\end{equation}

$N(-\omega)=\frac{1}{e^{-\beta \omega} -1}=-(1+\frac{1}{e^{-\beta \omega} -1})$, let $\omega'=-\omega$, in the meanwhile, we extend $J(\omega)$ to negative values of $\omega$ via  $J(-\omega)=-J(\omega)$.
\begin{equation}\begin{split}
&\int_{0}^{\infty} d\omega J(\omega) N(\omega)e^{i\omega(t+\chi)}\\
=&\int_{0}^{\infty} d(-\omega') J(-\omega') N(-\omega')e^{-i\omega'(t+\chi)}\\
=&-\int_{0}^{-\infty} d\omega' (-J(\omega')) (-(N(\omega')+1)) e^{-i\omega'(t+\chi)}\\
=&\int_{-\infty}^{0}d\omega'J(\omega')(N(\omega')+1)e^{-i\omega'(t+\chi)}\\
=&\int_{-\infty}^{0}d\omega J(\omega)(N(\omega)+1)e^{-i\omega(t+\chi)}
\end{split}
\end{equation}
Thus, $C(-2\chi,t)=\int_{-\infty}^{\infty}d\omega J(\omega)(N(\omega)+1)e^{-i\omega(t+\chi)}$. In the interaction picture, $\tilde{\pmb S}(t)=\pmb U^{\dag}(t,0)\pmb S\pmb U(t,0)$, where $\pmb U(t_1,t_0) = \vec{\pmb T}e^{-i\int_{t_0}^{t_1} d\tau \pmb H_S(\tau)}$. Replace them into the above equation and transform the equation back to Schr\"odinger's picture through $\pmb U(t,0)$ acting on the left side and $\pmb U^{\dag}(t,0)$ acting on the right side of the equation, we obtain

\begin{equation}
\begin{split}
 &i\pmb H_S(t)\pmb \rho(\chi, t)+\partial_t\pmb \rho(\chi, t) + \pmb \rho(\chi, t)(-i\pmb H_S(t))=\\
 &- \int^{\infty}_0 ds \lbrace  C(0,s) \pmb S\pmb U(t,t-s) \pmb S \pmb U^{\dag}(t,t-s) \pmb \rho(\chi, t)\\
 &- C(-2\chi,-s) \pmb S\pmb \rho(\chi, t)\pmb U(t,t-s) \pmb S \pmb U^{\dag}(t,t-s) \\
 &-  C(-2\chi,s)  \pmb U(t,t-s)\pmb S\pmb U^{\dag}(t,t-s)\pmb \rho(\chi, t) \pmb S\\
 &+  C(0,-s)  \pmb \rho(\chi, t) \pmb U(t,t-s)\pmb S \pmb U^{\dag}(t,t-s)\pmb S\rbrace
\end{split}
\end{equation}

By using Floquet theorem, $\pmb U(t,t_0)=\sum_{r}\exp(-i\varepsilon_r(t-t_0)) |r(t)\rangle\langle r(t_0)|$,

\begin{equation}
\begin{split}
 &\partial_t\pmb \rho(\chi, t)=-i[\pmb H_S(t), \pmb \rho(\chi, t)]\\
 &- \int^{\infty}_0 ds \lbrace  C(0,s) \pmb S \sum_{r}\exp(-i\varepsilon_r s) |r(t)\rangle\langle r(t-s)| \pmb S \sum_{r'}\exp(i\varepsilon_{r'} s) |r'(t-s)\rangle\langle r'(t)| \pmb \rho(\chi, t)\\
 &- C(-2\chi,-s) \pmb S\pmb \rho(\chi, t)\sum_{r}\exp(-i\varepsilon_r s) |r(t)\rangle\langle r(t-s)| \pmb S \sum_{r'}\exp(i\varepsilon_{r'} s) |r'(t-s)\rangle\langle r'(t)| \\
 &-  C(-2\chi,s) \sum_{r} \exp(-i\varepsilon_r s) |r(t)\rangle\langle r(t-s)|\pmb S\sum_{r'}\exp(i\varepsilon_{r'} s) |r'(t-s)\rangle\langle r'(t)|\pmb \rho(\chi, t) \pmb S\\
 &+  C(0,-s)  \pmb \rho(\chi, t) \sum_{r}\exp(-i\varepsilon_r s) |r(t)\rangle\langle r(t-s)|\pmb S \sum_{r'}\exp(i\varepsilon_{r'} s) |r'(t-s)\rangle\langle r'(t)|\pmb S\rbrace\\
 &=-i[\pmb H_S(t), \pmb \rho(\chi, t)]\\
&- \sum_{\omega}\int^{\infty}_0 ds \lbrace  C(0,s) \pmb S \exp(-i\omega s) \langle r(t-s)| \pmb S |r'(t-s)\rangle|r(t)\rangle\langle r'(t)| \pmb \rho(\chi, t)\\
 &- C(-2\chi,-s) \pmb S\pmb \rho(\chi, t)\exp(-i\omega s) \langle r(t-s)| \pmb S  |r'(t-s)\rangle|r(t)\rangle\langle r'(t)| \\
 &-  C(-2\chi,s) \exp(-i\omega s) \langle r(t-s)|\pmb S |r'(t-s)\rangle|r(t)\rangle\langle r'(t)|\pmb \rho(\chi, t) \pmb S\\
 &+  C(0,-s)  \pmb \rho(\chi, t) \exp(-i\omega s) \langle r(t-s)|\pmb S  |r'(t-s)\rangle|r(t)\rangle\langle r'(t)|\pmb S\rbrace
\end{split}
\end{equation}
$\langle r(t-s)|\pmb S |r'(t-s)\rangle$ is periodic with periodicity $T$ due to the periodicity of the Floquet mode $|r(t-s)\rangle$, we can take Fourier transform on it, 

\begin{equation}\begin{split}
 \pmb S_{\omega,n}&=\left[ \int_0^T \frac{d(t-s)}{T}  \langle r(t-s)| \pmb S \, e^{-i n \omega_L (t-s)} |r'(t-s)\rangle   \right]\, |r(t)\rangle \langle r'(t)|.
\end{split}
\end{equation}

Therefore, $\langle r(t-s)|\pmb S |r'(t-s)\rangle|r(t)\rangle \langle r'(t)|=\sum_{n} e^{i n \omega_L (t-s)} \pmb S_{\omega,n}$. Use $ C(-2\chi,t) = \frac{1}{\pi} \int^{\infty}_{-\infty} d \omega e^{- i \omega (\chi + t)} J(\omega) \left[ 1 + N_\omega \right] $ and $\langle r(t-s)|\pmb S |r'(t-s)\rangle|r(t)\rangle \langle r'(t)|=\sum_{n} e^{i n \omega_L (t-s)} \pmb S_{\omega,n}$, we arrive at
\begin{equation}
\begin{split}
 &\partial_t\pmb \rho(\chi, t)=-i[\pmb H_S(t), \pmb \rho(\chi, t)]\\
 &- \sum_{n,\omega} e^{i n \omega_l t} \int^{\infty}_0 ds \lbrace  C(0,s) \pmb S \exp(-i(\omega+n \omega_L) s) \pmb S_{\omega,n}\pmb \rho(\chi, t)\\
 &- C(-2\chi,-s) \pmb S\pmb \rho(\chi, t)\exp(-i(\omega+n \omega_L) s) \pmb S_{\omega,n} \\
 &-  C(-2\chi,s) \exp(-i(\omega+n \omega_L) s)\pmb S_{\omega,n}\pmb \rho(\chi, t) \pmb S\\
 &+  C(0,-s)  \pmb \rho(\chi, t) \exp(-i(\omega+n \omega_L) s) \pmb S_{\omega,n}\pmb S\rbrace\\
 &-i[\pmb H_S(t), \pmb \rho(\chi, t)]\\
 &=- \sum_{n,\omega}\lbrace  e^{i n \omega_L t} \frac{1}{\pi} \int^{\infty}_{-\infty} d \omega' \int^{\infty}_0 ds e^{- i (\omega'+(\omega+n \omega_L)) s} J(\omega') \left[ 1 + N_{\omega'} \right] \pmb S \pmb S_{\omega,n}\pmb \rho(\chi, t)\\
 &- \frac{1}{\pi} \int^{\infty}_{-\infty} d \omega' \int^{\infty}_0 ds e^{- i \omega' \chi }e^{i (\omega'-(\omega+n \omega_L))s} J(\omega') \left[ 1 + N_{\omega'} \right] \pmb S\pmb \rho(\chi, t)\pmb S_{\omega,n} \\
 &-  \frac{1}{\pi} \int^{\infty}_{-\infty} d \omega' \int^{\infty}_0 ds e^{- i \omega' \chi} e^{- i (\omega'+(\omega+n \omega_L)) s}J(\omega') \left[ 1 + N_{\omega'} \right] \pmb S_{\omega,n}\pmb \rho(\chi, t) \pmb S\\
 &+ \frac{1}{\pi} \int^{\infty}_{-\infty} d \omega' \int^{\infty}_0 ds e^{i (\omega'-(\omega+n \omega_L)) s} J(\omega') \left[ 1 + N_{\omega'} \right] \pmb \rho(\chi, t)\pmb S_{\omega,n}\pmb S\rbrace
\end{split}
\end{equation}

With the help of $\int^{\infty}_0  ds \, e^{i \omega s} = \pi \delta(\omega) + i \, \mathcal{P}  \, \frac{1}{\omega}$ and disregarding the principal value $\mathcal{P} $ term, $J(-\omega)=-J(\omega)$, and $1+N(-\omega)=1+\frac{1}{e^{-\beta\omega}-1}=\frac{1}{1-e^{\beta\omega}}=-N(\omega)$. We get the generalized master equation finally.

\begin{equation}
\begin{split}
&\partial_t\pmb \rho(\chi, t)=-i[\pmb H_S(t), \pmb \rho(\chi, t)]\\
&- \sum_{n,\omega} e^{i n \omega_l t} \lbrace J(\Delta_{\omega,n}) \left[N_{\Delta_{\omega,n}} \right] \pmb S \pmb S_{\omega,n}\pmb \rho(\chi, t)\\
 &- e^{-i\Delta_{\omega,n} \chi } J(\Delta_{\omega,n}) \left[ 1 + N_{\Delta_{\omega,n}} \right] \pmb S\pmb \rho(\chi, t)\pmb S_{\omega,n} \\
 &- e^{i \Delta_{\omega,n} \chi} J(\Delta_{\omega,n}) \left[N_{\Delta_{\omega,n}} \right] \pmb S_{\omega,n}\pmb \rho(\chi, t) \pmb S\\
 &+ J(\Delta_{\omega,n}) \left[ 1 + N_{\Delta_{\omega,n}} \right] \pmb \rho(\chi, t)\pmb S_{\omega,n}\pmb S\rbrace
\end{split}
\end{equation}

\section{Derivation of the effective system plus system-reservoirs interaction Hamiltonian through flow equation approach} \label{flow equation}

In this section, we derive the effective system plus system-reservoirs interaction Hamiltonian through flow equation approach \cite{Floquet3}. The method takes advantage of infinitesimal unitary transformation steps, from which renormalization-group–like flow equations are derived to derive the effective Hamiltonian. The flow equation is

\begin{equation}\label{exact flow}
\frac{d \pmb H(s,t)}{ds}=-\pmb V(s,t)+i\int_{0}^{t}dt_{1}[\pmb V(s,t_1),\pmb H(s,t)],
\end{equation}
where $s$ is the flow parameter, $\pmb H(s,t)$ and $\pmb V(s,t)$ are total Hamiltonian and its time-dependent part respectively. It should be noted that the family of Hamiltonians $\pmb H(s,t)$ represents an interpolation between a starting Hamiltonian $\pmb H(0,t)$ and a final Hamiltonian $\pmb H(\infty,t)$. Here, $\pmb H(\infty,t)$ is the Floquet Hamiltonian $\pmb H_F$ if $\pmb V(\infty,t)=0$. We set appropriate boundary conditions by enforcing that $s=0$ corresponds to the initial unchanged Hamiltonian.  $\pmb H(s,t)$ can be expressed as a sum of linear operators with coefficients $c_i(s,t)$, $\pmb H(s,t)=\sum_i c_i(s,t) \pmb O_i$. The $\pmb O_i$ operators are nothing other than kinetic and potential energy terms appearing in a Hamiltonian.  Note that the set of operators may include both the original operators and new terms generated from the commutator in Eqn.\ref{exact flow}.  The coefficients $c_i(s,t)$ describe the coupling constants (strength) of these terms. In this representation, Eqn.\ref{exact flow} can be written in a numerically tractable form,

\begin{equation}
\frac{d c_i(s,t)}{dt}=-g_i(t, [c_j(s,t')]), \quad\quad t' \in [0,T].
\end{equation}

In \cite{Floquet3}, Michael Vogl et al. put forward a more analytically tractable equation, which set $s=0$ only for the
terms $\pmb V(s,t)$. 

\begin{equation}
\frac{d \pmb H(s,t)}{ds}=-\pmb V(0,t)+i\int_{0}^{t}dt_{1}[\pmb V(0,t_1),\pmb H(s,t)],
\end{equation}
This corresponds to removing the original time-dependent part $\pmb V(s,t)$ from the Hamiltonian via the rotating frame transformation $e^{-i\int_{0}^{t}dt \pmb V(t)}$ while generating other new time dependences.
To ensure that this approximation actually corresponds to the aforementioned unitary transformation, we also need to restrict  $s \in [0, 1]$ rather than the previous $s \in [0, \infty)$. The effective time-independent Hamiltonian is then given by $\pmb H_{eff}= \sum_{i} \overline{c_i}(1,t) \pmb O_i$, where $\overline{c_i}(1,t)=\frac{1}{T}\int_{0}^{T}c_i(1,t)dt$, if we are only interested in stroboscopic dynamics.

We now derive the effective Hamiltonian for the model we studied. The system plus system-reservoirs interaction Hamiltonian is 

\begin{equation}\begin{split}
\pmb H_{S}(t)+\pmb H_{SB}=\frac{\omega_{A}+K\cos{(\omega_{L}t)}}{2}\pmb\sigma_{z}^{A}+\frac{\omega_{B}}{2}\pmb\sigma_{z}^{B}+\lambda(\pmb\sigma_{+}^{1}\pmb\sigma_{-}^{2}+\pmb\sigma_{-}^{1}\pmb\sigma_{+}^{2})+\pmb \sigma_{x}^{A} \pmb B_{A}+\pmb \sigma_{x}^{B} \pmb B_{B},
\end{split}
\end{equation}

where $\pmb B_i=\sum_K c_{ik}\pmb x_{ik}$ is reservoirs coupling operators and $\pmb V(0,t)=\frac{K\cos{(\omega_{L}t)}}{2}\pmb\sigma_{z}^{A}$ in this case. We make the ansatz

\begin{equation}
\pmb H(s,t)=C_{1}(s)\pmb\sigma_{z}^{A}+C_{2}(s)\pmb\sigma_{z}^{B}+C_{3}(s)\pmb\sigma_{+}^{1}\pmb\sigma_{-}^{2}+C_{4}(s)\pmb\sigma_{-}^{1}\pmb\sigma_{+}^{2}+C_{5}(s)\pmb \sigma_{x}^{A} \pmb B_{A}+C_{6}(s)\pmb \sigma_{x}^{B} \pmb B_{B}+C_{7}(s)\pmb \sigma_{y}^{A}\pmb B_{A},
\end{equation}

and then find the flow equation is 
\begin{equation}\begin{split}
\frac{d C_1(s)}{ds}&=-\frac{K\cos{(\omega_{L}t)}}{2}\\
\frac{d C_2(s)}{ds}&=0\\
\frac{d C_3(s)}{ds}&=\frac{iKC_{3}(s)}{\omega_{L}} \sin{(\omega_{L}t)}\\
\frac{d C_4(s)}{ds}&=-\frac{iKC_{4}(s)}{\omega_{L}} \sin{(\omega_{L}t)}\\
\frac{d C_5(s)}{ds}&=\frac{KC_{7}(s)}{\omega_{L}} \sin{(\omega_{L}t)}\\
\frac{d C_6(s)}{ds}&=0\\
\frac{d C_7(s)}{ds}&=-\frac{KC_{5}(s)}{\omega_{L}} \sin{(\omega_{L}t)}\\
\end{split}
\end{equation}
with the initial condition
\begin{equation}\begin{split}
 C_1(0)&=\frac{K\cos{(\omega_{L}t)}+\omega_A}{2}\\
 C_2(0)&=\frac{\omega_B}{2}\\
C_3(0)&=\lambda\\
C_4(0)&=\lambda\\
C_5(0)&=1\\
C_6(0)&=1\\
C_7(0)&=0\\
\end{split}
\end{equation}

Their solutions are 
\begin{equation}\begin{split}
C_1(s)&=\frac{1}{2} (\omega_A + K \cos(\omega_L t) - K s \cos(\omega_L t))\\
C_2(s)&=\frac{\omega_B}{2}\\
C_3(s)&=\lambda e^{\frac{i K s \sin(\omega_L t)}{\omega_L}}\\
C_4(s)&=\lambda e^{\frac{-i K s \sin(\omega_L t)}{\omega_L}}\\
C_5(s)&=\cos(K s \sin(\omega_L t)/\omega_L)\\
C_6(s)&=1\\
C_7(s)&=-\sin(K s \sin(\omega_L t)/\omega_L)\\
\end{split}
\end{equation}
After taking an average over one period and set $s=1$, we end up with the effective time-independent Hamiltonian
\begin{equation}\begin{split}
\pmb H_{eff}^{S}+\pmb H_{eff}^{SB}=\frac{\omega_{A}}{2}\pmb\sigma_{z}^{A}+\frac{\omega_{B}}{2}\pmb\sigma_{z}^{B}+\lambda J_0( K/\omega_L)(\pmb\sigma_{+}^{1}\pmb\sigma_{-}^{2}+\pmb\sigma_{-}^{1}\pmb\sigma_{+}^{2})+ J_0( K/\omega_L)\pmb \sigma_{x}^{A} \pmb B_{A}+\pmb \sigma_{x}^{B} \pmb B_{B},
\end{split}
\end{equation}

\end{document}